\documentclass[prl,aps,twocolumn,byrevtex,showpacs,floatfix,superscriptaddress,lengthcheck]{revtex4}

\usepackage{epsfig}

\begin{document}

\title{Phase-dependent decoherence of optical transitions in $\mathbf{Pr^{3+}:LaF_3}$ in the presence of a driving field}
\author{G. J. Pryde}
\email{pryde@physics.uq.edu.au}
\affiliation{Laser Physics Centre, Research School of Physical Sciences and Engineering, The Australian National University, Canberra, ACT, 0200, Australia}
\affiliation{Centre for Quantum Computer Technology, Department of Physics, The University of Queensland, Brisbane, 4072, Australia}
\author{M. J. Sellars}
\affiliation{Laser Physics Centre, Research School of Physical Sciences and Engineering, The Australian National University, Canberra, ACT, 0200, Australia}
\author{N. B. Manson}
\affiliation{Laser Physics Centre, Research School of Physical Sciences and Engineering, The Australian National University, Canberra, ACT, 0200, Australia}

\date{23 September 2003}

\begin{abstract}
The decoherence times of orthogonally phased components of the optical transition dipole moment in a two-level system have been observed to differ by an order of magnitude. This phase anisotropy is observed in coherent transient experiments where an optical driving field is present during extended periods of decoherence. The decoherence time of the component of the dipole moment in phase with the driving field is extended compared to $T_{2}$, obtained from two-pulse photon echoes, in analogy with the spin locking technique of NMR. This is the first phase-dependent investigation of optical decoherence in the presence of a driving field. 
\end{abstract}
\pacs{78.47.+p, 42.50.Hz, 42.50.Md}
\maketitle

There is strong and growing interest in the use of the coherence of optical transitions for quantum information and quantum technology applications. The realization of quantum computing with optical transitions is being pursued in a number of systems (Refs. \cite{sellarsqc,hemmerone,cirac,haroche}, for example); there have been dramatic demonstrations of slow light and light storage \cite{hemmertwo,walsworth,hau} and there also exists a well-established program to develop optical coherent processing techniques (Refs. \cite{cole,babbitt}, for example). Perhaps the most critical parameter in developing these technological ideas is the decoherence time of the transition dipole moment (TDM). Rare-earth doped crystals are especially interesting for these applications because of the long optical decoherence times that can be achieved. 

Although it is typical to quote a value for the decoherence time $T_{2}$ (as measured using photon echoes), decoherence in these optical impurity sites cannot adequately be described using a single parameter.  This was first demonstrated by DeVoe and Brewer \cite{devoe}, when the decoherence rate for the $^{3}$H$_{4}$ $\leftrightarrow$ $^{1}$D$_{2}$ optical transition of Pr$^{3+}$:LaF$_{3}$ in the presence of an optical driving field was found to be radically intensity dependent, in violation of the optical Bloch equations.  Similar effects have since been observed for a number of optical transitions \cite{szabo&muramoto,sellarsandmanson,mossberg}. Given that, in many systems, the applications listed above will require optically driving transitions for periods comparable to the decoherence time of the transition, it will be important to develop a detailed understanding of the effect of the driving field on the decoherence mechanisms. There is also a strong fundamental interest in forming a complete picture of decoherence in optical transitions generally.

Since both the optical driving field and TDM are quantities described by complex numbers, it is necessary to map out the intensity dependence as a function of their relative phase to fully characterize decoherence during interaction with the field. Until now, the possibility of a dependence of the dynamics on the relative phase of the driving field and the TDM has been experimentally neglected, despite the example of NMR (Ref. \cite{redfield}) and the suggestions of Sellars and Manson \cite{sellarsandmanson} and Shakhmuratov \cite{shakhmuratov}. We examine a well known system, Pr$^{3+}$:LaF$_{3}$ \cite{wald}, which like the majority of the transitions of interest in well ordered crystalline hosts at temperatures below 2 K, has a decoherence rate significantly in excess of that predicted from the lifetimes of the states. It is proposed that the extension of the observed decoherence time due to the driving field is independent of the specific decoherence mechanism, making these results applicable to any system where $T_{2}$ is not limited by $T_{1}$. 

\begin{figure}[!htb]
\center{\epsfig{figure=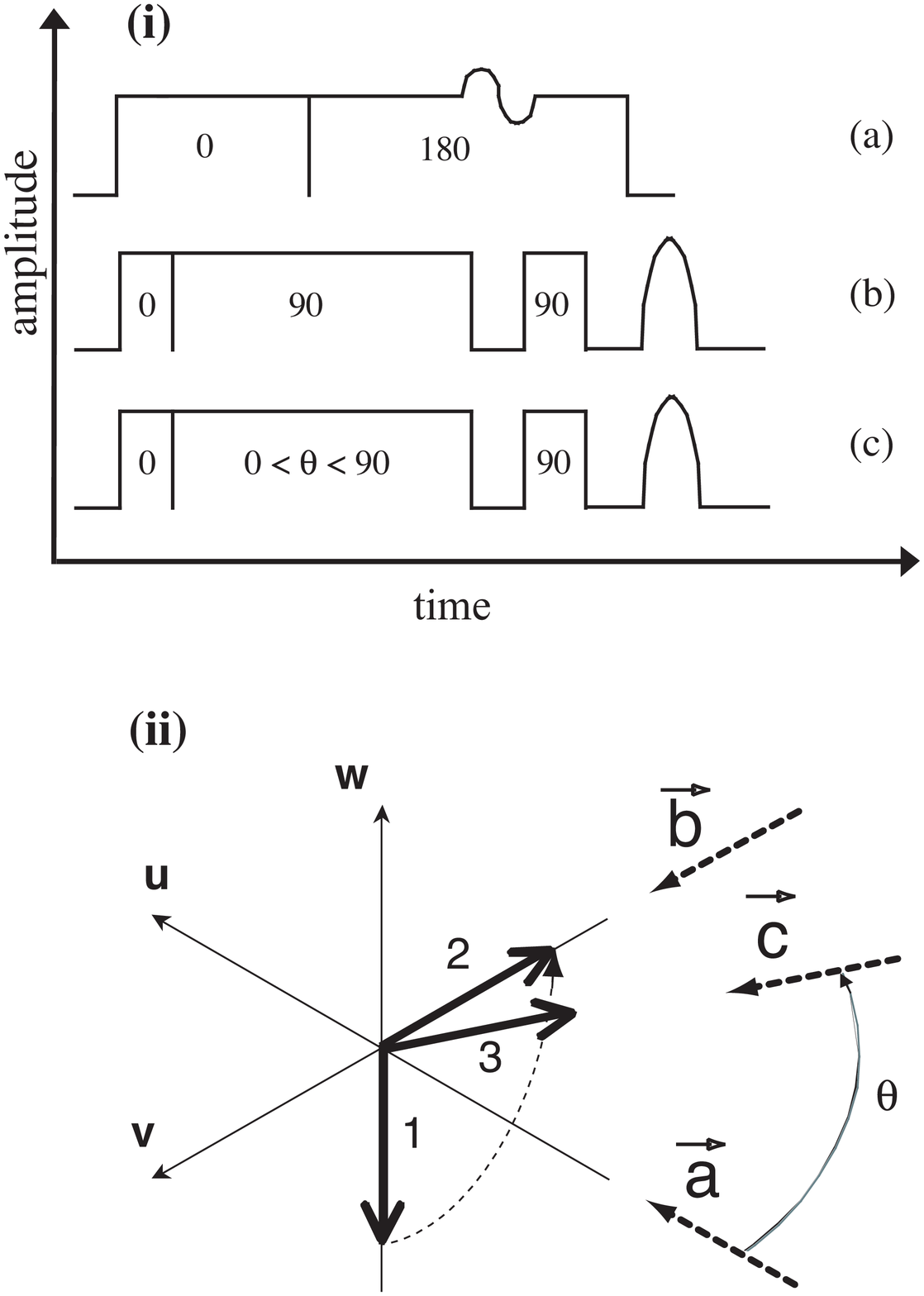,clip=true,width=\columnwidth,bbllx=100,bblly=140,bburx=705,bbury=920}}
\caption{(i)~(a) Rotary echo pulse sequence. (b,c) Radiation locking pulse sequences. The numbers represent the phase (degrees) of the optical field relative to the first pulse. The curved features schematically represent the coherent transient signals. (ii) Bloch picture of the phase relationships in the RE and RL experiments for the case of resonant ions. Angles in the uv plane represent phase. In the RE experiment, the optical field is parallel (antiparallel) to $\vec{a}$ during the $0^{\circ}\ (180^{\circ})$ pulse. The Bloch vector is driven from 1 to 2 and beyond, as the first pulse elapses, and is then reversed after the phase inversion of the optical field. The evolution of the Bloch vector is entirely in the vw plane, in quadrature with the driving field. The RE signal is observed for an ensemble when the Bloch vector is close to position 1, in quadrature with the driving field. For the RL, the Bloch vector is promoted from 1 to 2 by a $\pi /2$ pulse whose field vector lies along $\vec{a}\ (0^{\circ})$. If the second pulse has $90 ^{\circ}$ phase, as in (i)(b), a driving field with field vector $\vec{b}$ is applied during the decoherence time. The echo is read out with a $\pi$ pulse with phase $0^{\circ}$ ($\vec{a}$) or $90 ^{\circ}$ ($\vec{b}$), which are equivalent (to within a $180^{\circ}$ phase shift of the coherent transient signal). In the RL experiment, the Bloch vector is always lies along -v (position 2), in-phase (technically in-\emph{anti}phase, but as opposed to in-quadrature)  with the driving field, and for an ensemble, the echo arises from this component. The case described in (i)(c) is similar, except the driving field of phase $\theta$ now lies along $\vec{c}$, so that there exists an angle $90^{\circ}-\theta$ between the Bloch vector (which lies in position 2 at the commencement of the locking pulse) and the driving field. The echo signal measured is the component corresponding to the Bloch vector in position 3, the $\vec{c}$ direction.}
\label{fig:bloch1}
\end{figure}

To map out the dependence of the decoherence rates on the optical field, two optical coherent transient techniques were employed: rotary echoes (RE) and radiation locking (RL). Figure \ref{fig:bloch1} shows these pulse sequences and their effect in Bloch space. Two pulse photon echoes (2PE) were used to measure the decoherence time in the absence of a driving field, since the pulses in the 2PE sequence are brief compared to the evolution time of the system. The coherent transient experiments were conducted using a frequency-stabilized dye laser with a stability of  200 Hz over hundred-millisecond time scales and a Mach-Zehnder interferometer, with a frequency shift in one arm, to obtain phase sensitive heterodyne detection of the coherent transient signals \cite{pryde}. The crystal, immersed in a liquid He cryostat at 1.6 K, was oriented with the light propagating along the C$_{3}$ axis with a static magnetic field of 500 G perpendicular to C$_{3}$, to allow standardization with the experiments of DeVoe and Brewer \cite{devoe}. The interaction strength of the field and the transition (Rabi frequency) was controlled using the laser intensity.  

The RL pulse sequence \cite{sellarsandmanson} consists of a $\pi /2$ pulse followed by a long (locking) pulse, shifted in phase (see Fig.~\ref{fig:bloch1}). Although it was possible to observe the coherence as an FID at the end of the locking pulse, switching transients tended to distort the signal. To avoid this problem, a $\pi$ pulse was used to rephase this coherence of the TDM and observe it as an echo. It is normal to use a phase shift of $90 ^{\circ}$ between the $\pi /2$ and locking pulses in the RL sequence. The resonant ions are promoted to an equal superposition of the ground and excited states by the first pulse, represented by their Bloch vectors lying along the $v$ axis. The locking field then holds the Bloch vector along $v$, analogous to spin locking in NMR \cite{solomon}. Bloch vectors corresponding to nonresonant ions do not lie strictly along $v$ after the $\pi /2$ pulse and they precess in a cone about the locking field. During the locking period, the component of the off-resonant Bloch vectors in-quadrature with the driving field dephases, but the in-phase component is locked in the same way as for resonant ions. The coherence observed in the echo is from the component of the TDM in phase with the driving field, the other component having dephased. An artifact is observed in the coherent transient signal due to the inhomogeneous broadening - the locking and rephasing pulses together contribute a two pulse ``echo'' that appears entirely in the quadrature component. This artifact is easily separated from the real signal with phase sensitive detection.   

In the current work, phase shifts (between $\pi /2$ and locking pulses) other than $90 ^{\circ}$ were also employed. In this case, even the resonant ions have Bloch vectors with some component in-quadrature with the driving field. By varying the angle, $\theta$, between the TDM and the driving field, we could test for a dependence of the decoherence on $\theta$. Regardless of $\theta$, the real signal observed corresponds to the component of the TDM in-phase with the driving field.    

The RE pulse sequence \cite{wong} consists of two oppositely-phased pulses as shown in Fig.~\ref{fig:bloch1}. Bloch vectors corresponding to resonant ions are rotated in the $vw$ plane by the first pulse. The phase-inverted pulse also achieves rotation in the $vw$ plane, but in the opposite sense. An echo is observed when the second pulse has rotated the Bloch vectors back to their original state and rephases them. Bloch vectors corresponding to non-resonant ions have non-zero $u$ components during the pulses, but these average to zero for a frequency-symmetric ensemble of ions. The RE was used to observe the effect of having the TDM in-quadrature with the driving field, since the ensemble-average Bloch vector is always in-quadrature.  

\begin{figure}[!htb]
\center{\epsfig{figure=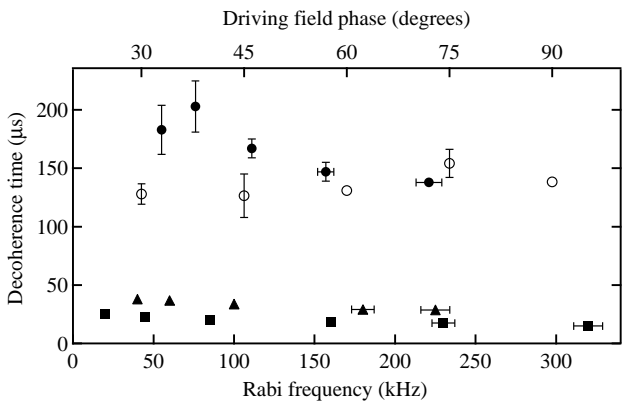,clip=true,width=\columnwidth,bbllx=15,bblly=10,bburx=690,bbury=420}}
\caption{Measured decoherence times as a function of Rabi frequency (filled markers) for the different coherent transient pulse sequences: RL (circles), RE (triangles), 2PE (squares); and as a function of driving field phase (open markers) for the RL experiments. Error bars not shown are smaller than the markers.}
\label{fig:rlre2pe}
\end{figure}

The measurements of decoherence during driving, observed in the coherent transient experiments and presented in Fig.~\ref{fig:rlre2pe}, can be easily summarized. The 2PE experiments define $T_{2}$ in the limit of zero Rabi frequency, and the data shows that they produce a decoherence time that is approximately equal to $T_{2}$ for all Rabi frequencies. The RE experiments yield a decoherence time that is approximately equal to $2T_{2}$ for all Rabi frequencies, as predicted by the optical Bloch equations (OBEs). The RL experiments yield decoherence times that are not predicted by the OBEs and are extended by up to a factor of 9.2 relative to $T_{2}$, demonstrating a large suppression of decoherence in the TDM component in-phase with the driving field. This is a clear demonstration that the orthogonally-phased components of the TDM decohere at different rates in the presence of a driving field.     

The dependence of the decoherence of the in-phase component of the RL as the phase angle of the driving field is varied is also shown in Fig.~\ref{fig:rlre2pe}. Here, the Rabi frequency of the driving field is held constant at 225 kHz. It can be seen that the dependence of the decoherence time of the in-phase component is nearly independent of the phase of the driving field with respect to the TDM.

The combined result that can be derived from Fig.~\ref{fig:rlre2pe} is that for a given Rabi frequency, the decoherence time of the in-phase component of the TDM is extended uniformly regardless of the relative phase of the TDM and driving field, while the in-quadrature component of the TDM decoheres, in the case of in-quadrature driving, at the expected rate based on 2PE measurements. Unfortunately, the artefacts due to inhomogeneous broadening prevent making a statement on the decoherence of the in-quadrature component for driving fields not in-quadrature with the TDM . 

The elongation of the decoherence time of the in-phase component can be attributed to radiation locking, the optical analogy of spin locking in NMR. The term {\em locking} refers to the fact that the driving field maintains, or locks, the in-phase component of the Bloch vector at a fixed phase (the projection of the Bloch vector onto the coherence plane $uv$ represents the TDM in the optical experiment.) Fluctuations in the transition frequency that tend to disturb this phase are averaged out by the locking field when $\chi/2\pi > \delta$, where $\chi/2\pi$ is the Rabi frequency and $\delta$ is the amplitude of the fluctuations \cite{redfield,shakhmuratov}. The component of the TDM that is in-quadrature with the driving field is expected to be unaffected by the presence of the driving field - no locking takes place because of the orthogonality. The observed results correlate with the results of Redfield, who observed the same phase anisotropy for NMR saturation experiments performed on metals \cite{redfield}. In Pr$^{3+}$:LaF$_{3}$ at low temperature, $\delta \approx$ 8-10 kHz \cite{kessel}. This interpretation is clearly consistent with the RL results presented here. It was not possible to use the RL sequence with Rabi frequencies at the 10 kHz level or below, because of signal-to-noise concerns (particular to the present experimental realization, and not generally limiting at this level). However, the turn-on of radiation locking effects at approximately this Rabi frequency is observed in FID experiments \cite{devoe}, where the decay time begins to deviate from the OBE prediction at $\chi/2\pi \approx$4 kHz. Sellars and Manson have explained that, in practice, saturation causes the driving field in a FID experiment to be largely in-phase with the TDM, as in RL experiments \cite{sellarsandmanson}. 

Inhomogeneous broadening plays a significant, although not dominant, part in all of the coherent transient experiments reported. The 2PE and RE are least affected by the inhomogeneous broadening, since the concept of an echo is to completely reverse the dephasing due to inhomogeneous broadening, leaving only effects independent of the inhomogeneous broadening as contributors to the measured decoherence time. The main disadvantage of the inhomogeneous broadening in the present measurements is that it complicates the interpretation of RL experiments. The driving field and Bloch vectors corresponding to ions of all frequencies are at an angle $90^{\circ}-\theta$ in the ideal experiment, leading to output coherent transient components in both phases. In principle, one could extract the in-quadrature component of the coherent transient signal by inverting the phase of the driving field mid-pulse, and observing a rotary echo in-quadrature as well as the in-phase RL signals.  This would lead to simultaneous decoherence time measurements for the two phases. However, in the real RL experiment, we observe that the effects of inhomogeneous broadening dominate the in-quadrature signal, masking the in-quadrature signal due to resonantly excited ions (off-resonant ions make no net contribution to the in-phase signal.)  Furthermore, simple calculations, based on the assumptions of our interpretation of the lengthened in-phase decoherence time, indicate that the high power rolloff of the decoherence time, shown in fig. \ref{fig:rlre2pe} (filled circles), for the RL is an a artefact of inhomogeneous broadening. Despite the complications of inhomogeneous broadening, the experiments presented here clearly demonstrate the anisotropy of the decoherence of the two orthogonal components of the TDM in the presence of a driving field, and the independence of the decoherence time of the in-phase component of the TDM from the relative angle of the TDM and the driving field. This latter conclusion supports the hypothesis that it is reasonable to resolve the TDM into its in-phase and in-quadrature components and think of the driving field acting on these components separately. 

It is important to note that neither the phase dependence of decoherence during driving, nor the effects of inhomogeneous broadening, are peculiar to impurity-ion solids. The arguments also apply to atomic vapour spectroscopy where collisional dephasing and Doppler broadening are the decoherence and inhomogeneous broadening mechanisms respectively, and to other solid state systems such as quantum dots. In this respect, the present phase dependent study yields important general results. Although the simple radiation locking picture gives an adequate description of the extension of one  component of $T_2$ in many systems, a detailed description of the system-dependent decoherence effects under general conditions may require a more complicated approach. The example of Pr$^{3+}$:LaF$_{3}$ is a case in point.

A series of stochastic models \cite{kessel} have been proposed to account for the previous FID and rotary echo studies in Pr$^{3+}$:LaF$_{3}$ and ruby \cite{devoe,szabo&muramoto}.  The starting point for all these models is the assumption that the system can be treated as an optically active two-level ion perturbed by a time dependent field which, it turn, is generated by a bath composed of the coupled spins in the host. These models do not explain consistently the complete set of observed experimental results, possibly because of the neglect of the phase dependence of the decoherence on the driving field.  Kessel' et al. have produced a model \cite{kessel} that qualitatively agrees with the Redfield result and also with the work presented here and includes the idea of different decoherence times for the two components of the TDM. However, all the various models should be treated with caution since, in practice, the interaction between the optically active ion and the nearby host spins is stronger than the interactions between the host spins, making the ion/bath approximation invalid. 

The typical interaction strength between the Pr and the nearest neighbour F is of the order 10 kHz \cite{wald}, whilst the interaction between neighboring Fluorines is in the range 1 - 10 kHz \cite{muramoto}. These relative interaction strengths allow the possibility of driving one or more Fluorine nuclear spin flips whilst optically driving the $^{3}$H$_{4}$ $\leftrightarrow$ $^{1}$D$_{2}$ transition. We have calculated the approximate energy level structure and transition probabilities arising from the coupling of the Pr$^{3+}$ ion to the nuclei of five of the nearest-neighbour fluorines in the lattice and observed that there are multiple ($> 15$) coherence pathways (i.e. optical transitions from a manifold of superhyperfine levels to another such manifold) with transition strengths on the same order. This result is not strongly dependent on which five nearest neighbour nuclei are chosen. This calculation would be enhanced by inclusion of many more F neighbours, but it gives an indication of the importance of the superhyperfine interaction. In addition, hole burning experiments have displayed side hole structure arising from the superhyperfine interaction, showing up to 20\% of the transition strength associated with the flipping of nearest neighbour F nuclei \cite{sellarsthesis}.  It would therefore seem necessary that any theory of optical decoherence in Pr$^{3+}$:LaF$_{3}$ must incorporate optical excitation from one superhyperfine manifold to another. It would also be possible to address the description of decoherence in these systems from an experimental perspective, by repeating the experiments presented in this paper in systems where the interactions within the bath are much stronger than the optically active ion's interaction with the bath, and vice versa. Examples of appropriate rare-earth doped solids would be Eu$^{3+}$:LaF$_{3}$ for the former case and Pr$^{3+}$:Y$_{2}$SiO$_{5}$ for the latter, since Eu$^{3+}$-F interactions are much weaker than Pr$^{3+}$-F interactions and the Y-Y nuclear interaction is much weaker than that for F-F. Experimental data from two systems such as this, coupled with the  Pr$^{3+}$:LaF$_{3}$ data presented here, would form the basis for a detailed theory of optical decoherence due to magnetic couplings in solids; a theory that did not rely on the approximation of a two-level ion interacting with a bath.

In conclusion, the results presented form the first explicit phase-dependent study of optical decoherence in the presence of a driving field. The decoherence time of the component of the TDM in phase with a driving field has shown to be extended by nearly a factor of 10 relative to the photon echo measurement. The decoherence time of the TDM in quadrature with the driving field is not extended. These results are unaffected by the relative phase of the driving field and the prepared phase of the resonant ions. In analogy with spin locking in NMR, these results are interpreted as the rapid averaging out of the perturbations limiting the $T_{2}$ measured by photon echo or rotary echo experiments. This interpretation leads to a consistent explanation of the observed optical non-Bloch behaviour in Pr$^{3+}$:LaF$_{3}$ and ruby. The results will play an important part in the construction of quantum information demonstrations using optical transitions, and the fundamental understanding of the decoherence of transition dipole moments.


\begin{thebibliography}{22}
\expandafter\ifx\csname natexlab\endcsname\relax\def\natexlab#1{#1}\fi
\expandafter\ifx\csname bibnamefont\endcsname\relax
  \def\bibnamefont#1{#1}\fi
\expandafter\ifx\csname bibfnamefont\endcsname\relax
  \def\bibfnamefont#1{#1}\fi
\expandafter\ifx\csname citenamefont\endcsname\relax
  \def\citenamefont#1{#1}\fi
\expandafter\ifx\csname url\endcsname\relax
  \def\url#1{\texttt{#1}}\fi
\expandafter\ifx\csname urlprefix\endcsname\relax\def\urlprefix{URL }\fi
\providecommand{\bibinfo}[2]{#2}
\providecommand{\eprint}[2][]{\url{#2}}

\bibitem[{\citenamefont{Longdell and Sellars}(2002)}]{sellarsqc}
\bibinfo{author}{\bibfnamefont{J.~J.} \bibnamefont{Longdell}} \bibnamefont{and}
  \bibinfo{author}{\bibfnamefont{M.~J.} \bibnamefont{Sellars}},
  \bibinfo{journal}{quant-ph/0310105}, Phys. Rev. A (to be published); \bibinfo{author}{\bibfnamefont{P.~R.} \bibnamefont{Hemmer}},
  \bibinfo{author}{\bibfnamefont{A.~V.} \bibnamefont{Turukhin}},
  \bibinfo{author}{\bibfnamefont{M.~S.} \bibnamefont{Shahriar}}
  \bibnamefont{and} \bibinfo{author}{\bibfnamefont{J.~A.}
  \bibnamefont{Musser}}, \bibinfo{journal}{Opt. Lett.}
  \textbf{\bibinfo{volume}{26}}, \bibinfo{pages}{361} (\bibinfo{year}{2001}).

\bibitem[{\citenamefont{Hemmer et~al.}(2001)\citenamefont{Hemmer, Turukhin,
  Shahriar, and Musser}}]{hemmerone}
\bibinfo{author}{\bibfnamefont{F.} \bibnamefont{Schmidt-Kaler}},
  \bibinfo{author}{\bibfnamefont{H.} \bibnamefont{Haffner}},
  \bibinfo{author}{\bibfnamefont{M.} \bibnamefont{Riebe}},
   \bibinfo{author}{\bibfnamefont{S.} \bibnamefont{Gulde}},
    \bibinfo{author}{\bibfnamefont{G.~P.~T.} \bibnamefont{Lancaster}},
     \bibinfo{author}{\bibfnamefont{T.} \bibnamefont{Deuschle}},
      \bibinfo{author}{\bibfnamefont{C.} \bibnamefont{Becher}},
       \bibinfo{author}{\bibfnamefont{C.~F.} \bibnamefont{Roos}},
        \bibinfo{author}{\bibfnamefont{J.} \bibnamefont{Eschner}}
  \bibnamefont{and} \bibinfo{author}{\bibfnamefont{R.}
  \bibnamefont{Blatt}}, \bibinfo{journal}{Nature (London)}
  \textbf{\bibinfo{volume}{422}}, \bibinfo{pages}{408} (\bibinfo{year}{2003});
  \bibinfo{author}{\bibfnamefont{D.} \bibnamefont{Leibfried}},
  \bibinfo{author}{\bibfnamefont{B.} \bibnamefont{DeMarco}},
  \bibinfo{author}{\bibfnamefont{V.} \bibnamefont{Meyer}},
   \bibinfo{author}{\bibfnamefont{D.} \bibnamefont{Lucas}},
    \bibinfo{author}{\bibfnamefont{M.} \bibnamefont{Barrett}},
     \bibinfo{author}{\bibfnamefont{J.} \bibnamefont{Britton}},
      \bibinfo{author}{\bibfnamefont{W.~M.} \bibnamefont{Itano}},
       \bibinfo{author}{\bibfnamefont{B.} \bibnamefont{Jelenkovic}},
        \bibinfo{author}{\bibfnamefont{C.} \bibnamefont{Langer}},
         \bibinfo{author}{\bibfnamefont{T.} \bibnamefont{Rosenband}}
  \bibnamefont{and} \bibinfo{author}{\bibfnamefont{D.~J.}
  \bibnamefont{Wineland}}, \bibinfo{journal}{{\em ibid.}}
  \textbf{\bibinfo{volume}{422}}, \bibinfo{pages}{412} (\bibinfo{year}{2003});

\bibitem[{\citenamefont{Cirac and Zoller}(1995)}]{cirac}
\bibinfo{author}{\bibfnamefont{X.} \bibnamefont{Li}},
\bibinfo{author}{\bibfnamefont{Y.} \bibnamefont{Wu}},
\bibinfo{author}{\bibfnamefont{D.} \bibnamefont{Steel}},
\bibinfo{author}{\bibfnamefont{D.} \bibnamefont{Gammon}},
\bibinfo{author}{\bibfnamefont{T.~H.} \bibnamefont{Stievater}},
\bibinfo{author}{\bibfnamefont{D.~S.} \bibnamefont{Katzer}},
\bibinfo{author}{\bibfnamefont{D.} \bibnamefont{Park}},
\bibinfo{author}{\bibfnamefont{C.} \bibnamefont{Piermarocchi}}
 \bibnamefont{and}
  \bibinfo{author}{\bibfnamefont{L.~J.}~\bibnamefont{Sham}},
  \bibinfo{journal}{Science} \textbf{\bibinfo{volume}{301}},
  \bibinfo{pages}{809} (\bibinfo{year}{2003}).

\bibitem[{\citenamefont{Domokos et~al.}(1995)\citenamefont{Domokos, Raimond,
  Brune, and Haroche}}]{haroche}
\bibinfo{author}{\bibfnamefont{P.}~\bibnamefont{Domokos}},
  \bibinfo{author}{\bibfnamefont{J.~M.} \bibnamefont{Raimond}},
  \bibinfo{author}{\bibfnamefont{M.}~\bibnamefont{Brune}}, \bibnamefont{and}
  \bibinfo{author}{\bibfnamefont{S.}~\bibnamefont{Haroche}},
  \bibinfo{journal}{Phys. Rev. A} \textbf{\bibinfo{volume}{52}},
  \bibinfo{pages}{3554} (\bibinfo{year}{1995}).

\bibitem[{\citenamefont{Turukhin et~al.}(2002)\citenamefont{Turukhin,
  Sudarshanam, Shahriar, Musser, Ham, and Hemmer}}]{hemmertwo}
\bibinfo{author}{\bibfnamefont{A.~V.} \bibnamefont{Turukhin}},
  \bibinfo{author}{\bibfnamefont{V.~S.} \bibnamefont{Sudarshanam}},
  \bibinfo{author}{\bibfnamefont{M.~S.} \bibnamefont{Shahriar}},
  \bibinfo{author}{\bibfnamefont{J.~A.} \bibnamefont{Musser}},
  \bibinfo{author}{\bibfnamefont{B.~S.} \bibnamefont{Ham}}, \bibnamefont{and}
  \bibinfo{author}{\bibfnamefont{P.~R.} \bibnamefont{Hemmer}},
  \bibinfo{journal}{Phys. Rev. Lett.} \textbf{\bibinfo{volume}{88}},
  \bibinfo{pages}{023602} (\bibinfo{year}{2002}).

\bibitem[{\citenamefont{Phillips et~al.}(2001)\citenamefont{Phillips,
  Fleischhauer, Mair, Walsworth, and Lukin}}]{walsworth}
\bibinfo{author}{\bibfnamefont{D.~F.} \bibnamefont{Phillips}},
  \bibinfo{author}{\bibfnamefont{A.}~\bibnamefont{Fleischhauer}},
  \bibinfo{author}{\bibfnamefont{A.}~\bibnamefont{Mair}},
  \bibinfo{author}{\bibfnamefont{R.~L.} \bibnamefont{Walsworth}},
  \bibnamefont{and} \bibinfo{author}{\bibfnamefont{M.~D.} \bibnamefont{Lukin}},
  \bibinfo{journal}{Phys. Rev. Lett.} \textbf{\bibinfo{volume}{86}},
  \bibinfo{pages}{783} (\bibinfo{year}{2001}).

\bibitem[{\citenamefont{Liu et~al.}(2001)\citenamefont{Liu, Dutton, Behroozi,
  and Hau}}]{hau}
\bibinfo{author}{\bibfnamefont{C.}~\bibnamefont{Liu}},
  \bibinfo{author}{\bibfnamefont{Z.}~\bibnamefont{Dutton}},
  \bibinfo{author}{\bibfnamefont{C.~H.} \bibnamefont{Behroozi}},
  \bibnamefont{and} \bibinfo{author}{\bibfnamefont{L.~V.} \bibnamefont{Hau}},
  \bibinfo{journal}{Nature} \textbf{\bibinfo{volume}{409}},
  \bibinfo{pages}{490} (\bibinfo{year}{2001}).

\bibitem[{\citenamefont{Tian et~al.}(2001)\citenamefont{Tian, Reibel, and
  Babbitt}}]{babbitt}
\bibinfo{author}{\bibfnamefont{M.~Z.} \bibnamefont{Tian}},
  \bibinfo{author}{\bibfnamefont{R.}~\bibnamefont{Reibel}}, \bibnamefont{and}
  \bibinfo{author}{\bibfnamefont{W.~R.} \bibnamefont{Babbitt}},
  \bibinfo{journal}{Opt. Lett.} \textbf{\bibinfo{volume}{26}},
  \bibinfo{pages}{1143} (\bibinfo{year}{2001}).

\bibitem[{\citenamefont{Cole et~al.}(2002)\citenamefont{Cole, Bottger, Mohan,
  Reibel, Babbitt, Cone, and Merkel}}]{cole}
\bibinfo{author}{\bibfnamefont{Z.}~\bibnamefont{Cole}},
  \bibinfo{author}{\bibfnamefont{T.}~\bibnamefont{Bottger}},
  \bibinfo{author}{\bibfnamefont{R.~K.} \bibnamefont{Mohan}},
  \bibinfo{author}{\bibfnamefont{R.}~\bibnamefont{Reibel}},
  \bibinfo{author}{\bibfnamefont{W.~R.} \bibnamefont{Babbitt}},
  \bibinfo{author}{\bibfnamefont{R.~L.} \bibnamefont{Cone}}, \bibnamefont{and}
  \bibinfo{author}{\bibfnamefont{K.~D.} \bibnamefont{Merkel}},
  \bibinfo{journal}{Appl. Phys. Lett.} \textbf{\bibinfo{volume}{81}},
  \bibinfo{pages}{3525} (\bibinfo{year}{2002}).

\bibitem[{\citenamefont{DeVoe and Brewer}(1983)}]{devoe}
\bibinfo{author}{\bibfnamefont{R.~G.} \bibnamefont{DeVoe}} \bibnamefont{and}
  \bibinfo{author}{\bibfnamefont{R.~G.} \bibnamefont{Brewer}},
  \bibinfo{journal}{Phys. Rev. Lett.} \textbf{\bibinfo{volume}{50}},
  \bibinfo{pages}{1269} (\bibinfo{year}{1983}).

\bibitem[{\citenamefont{Szabo and Muramoto}(1989)}]{szabo&muramoto}
\bibinfo{author}{\bibfnamefont{A.}~\bibnamefont{Szabo}} \bibnamefont{and}
  \bibinfo{author}{\bibfnamefont{T.}~\bibnamefont{Muramoto}},
  \bibinfo{journal}{Phys. Rev. A} \textbf{\bibinfo{volume}{39}},
  \bibinfo{pages}{3992} (\bibinfo{year}{1989}).

\bibitem[{\citenamefont{Sellars and Manson}(1998)}]{sellarsandmanson}
\bibinfo{author}{\bibfnamefont{M.~J.} \bibnamefont{Sellars}} \bibnamefont{and}
  \bibinfo{author}{\bibfnamefont{N.~B.} \bibnamefont{Manson}},
  \bibinfo{journal}{J. Lumin.} \textbf{\bibinfo{volume}{76\&77}},
  \bibinfo{pages}{137} (\bibinfo{year}{1998}).

\bibitem[{\citenamefont{Yodh et~al.}(1984)\citenamefont{Yodh, Golub, Carlson,
  and Mossberg}}]{mossberg}
\bibinfo{author}{\bibfnamefont{A.~G.} \bibnamefont{Yodh}},
  \bibinfo{author}{\bibfnamefont{J.}~\bibnamefont{Golub}},
  \bibinfo{author}{\bibfnamefont{N.~W.} \bibnamefont{Carlson}},
  \bibnamefont{and} \bibinfo{author}{\bibfnamefont{T.~W.}
  \bibnamefont{Mossberg}}, \bibinfo{journal}{Phys. Rev. Lett.}
  \textbf{\bibinfo{volume}{53}}, \bibinfo{pages}{659} (\bibinfo{year}{1984}).

\bibitem[{\citenamefont{Redfield}(1955)}]{redfield}
\bibinfo{author}{\bibfnamefont{A.~G.} \bibnamefont{Redfield}},
  \bibinfo{journal}{Phys. Rev.} \textbf{\bibinfo{volume}{98}},
  \bibinfo{pages}{1787} (\bibinfo{year}{1955}).

\bibitem[{\citenamefont{Shakhmuratov}(1999)}]{shakhmuratov}
\bibinfo{author}{\bibfnamefont{R.~N.} \bibnamefont{Shakhmuratov}},
  \bibinfo{journal}{Phys. Rev. A} \textbf{\bibinfo{volume}{59}},
  \bibinfo{pages}{3788} (\bibinfo{year}{1999}).

\bibitem[{\citenamefont{Wald et~al.}(1992)\citenamefont{Wald, Hahn, and
  Lukac}}]{wald}
\bibinfo{author}{\bibfnamefont{L.~L.} \bibnamefont{Wald}},
  \bibinfo{author}{\bibfnamefont{E.~L.} \bibnamefont{Hahn}}, \bibnamefont{and}
  \bibinfo{author}{\bibfnamefont{M.}~\bibnamefont{Lukac}}, \bibinfo{journal}{J.
  Opt. Soc. Am. B} \textbf{\bibinfo{volume}{9}}, \bibinfo{pages}{789}
  (\bibinfo{year}{1992}), \bibinfo{note}{and references therein.}

\bibitem[{\citenamefont{Pryde et~al.}(2000)\citenamefont{Pryde, Sellars, and
  Manson}}]{pryde}
\bibinfo{author}{\bibfnamefont{G.~J.} \bibnamefont{Pryde}},
  \bibinfo{author}{\bibfnamefont{M.~J.} \bibnamefont{Sellars}},
  \bibnamefont{and} \bibinfo{author}{\bibfnamefont{N.~B.}
  \bibnamefont{Manson}}, \bibinfo{journal}{Phys. Rev. Lett.}
  \textbf{\bibinfo{volume}{84}}, \bibinfo{pages}{1152} (\bibinfo{year}{2000}).

\bibitem[{\citenamefont{Solomon}(1959)}]{solomon}
\bibinfo{author}{\bibfnamefont{I.}~\bibnamefont{Solomon}}, \bibinfo{journal}{C.
  R. Hebd Seances Acad. Sci.} \textbf{\bibinfo{volume}{248}},
  \bibinfo{pages}{92} (\bibinfo{year}{1959}).

\bibitem[{\citenamefont{Wong et~al.}(1980)\citenamefont{Wong, Kano, and
  Brewer}}]{wong}
\bibinfo{author}{\bibfnamefont{N.~C.} \bibnamefont{Wong}},
  \bibinfo{author}{\bibfnamefont{S.~S.} \bibnamefont{Kano}}, \bibnamefont{and}
  \bibinfo{author}{\bibfnamefont{R.~G.} \bibnamefont{Brewer}},
  \bibinfo{journal}{Phys. Rev. A} \textbf{\bibinfo{volume}{21}},
  \bibinfo{pages}{260} (\bibinfo{year}{1980}).

\bibitem[{\citenamefont{Kessel' et~al.}(1988)\citenamefont{Kessel',
  Shakhmuratov, and Eksin}}]{kessel}
\bibinfo{author}{\bibfnamefont{A.~R.} \bibnamefont{Kessel'}},
  \bibinfo{author}{\bibfnamefont{R.~P.} \bibnamefont{Shakhmuratov}},
  \bibnamefont{and} \bibinfo{author}{\bibfnamefont{L.~D.} \bibnamefont{Eksin}},
  \bibinfo{journal}{Zh. Eksp. Teor. Fiz.} \textbf{\bibinfo{volume}{94}},
  \bibinfo{pages}{202} (\bibinfo{year}{1988}), \bibinfo{note}{[Sov. Phys. JETP
  {\bf 67}, 2071 (1988)]}, and references therein.

\bibitem[{\citenamefont{Muramoto et~al.}(1992)\citenamefont{Muramoto,
  Takahashi, and Hashi}}]{muramoto}
\bibinfo{author}{\bibfnamefont{T.}~\bibnamefont{Muramoto}},
  \bibinfo{author}{\bibfnamefont{Y.}~\bibnamefont{Takahashi}},
  \bibnamefont{and} \bibinfo{author}{\bibfnamefont{T.}~\bibnamefont{Hashi}},
  \bibinfo{journal}{J. Lumin.} \textbf{\bibinfo{volume}{53}},
  \bibinfo{pages}{84} (\bibinfo{year}{1992}).

\bibitem[{\citenamefont{Sellars}(1995)}]{sellarsthesis}
\bibinfo{author}{\bibfnamefont{M.~J.} \bibnamefont{Sellars}},
  \bibinfo{title}{Ph.D. thesis}, \bibinfo{publisher}{The Australian
  National University}, \bibinfo{year}{1995}.

\end{thebibliography}
\end{document}